\documentclass{vgtc}                          

\ifpdf
  \pdfoutput=1\relax                   
  \usepackage{graphicx}                
  \DeclareGraphicsExtensions{.pdf,.png,.jpg,.jpeg} 
\else
  \usepackage{graphicx}                
  \DeclareGraphicsExtensions{.eps}     
\fi%

\graphicspath{{figures/}{pictures/}{images/}{./}} 

\usepackage{microtype}                 
\PassOptionsToPackage{warn}{textcomp}  
\usepackage{textcomp}                  
\usepackage{mathptmx}                  
\usepackage{times}                     
\usepackage{cite}                      
\usepackage{tabu}                      
\usepackage{booktabs}                  
\usepackage{enumitem}

\onlineid{0}

\vgtccategory{Research}

\vgtcinsertpkg

\title{AI Agents and the Future of VIS}

\author{Chen Zhu-Tian\thanks{with U of Minnesota, \{ztchen, pan00342, qianwen\}@umn.edu},
Nam Wook Kim\thanks{with Boston College, nam.wook.kim@bc.edu},
Saeed Boorboor\thanks{with UIC, boorboor@uic.edu},
Shivam Raval\thanks{with Harvard, sraval@g.harvard.edu},
Pan Hao\footnotemark[1]
Qianwen Wang\footnotemark[1],
Vidya Setlur\thanks{with Tableau Research, vsetlur@salesforce.com}
}

\abstract{Recent advances in agents (i.e., autonomous, goal-driven AI systems that iteratively observe, act, and learn from their environments) offer a fundamentally different approach from traditional AI models that passively respond to input. These AI agents are rapidly reshaping how we approach data-intensive tasks and providing new opportunities for the VIS community. Imagine an agent autonomously generating visualizations to analyze complex data, discovering patterns collaboratively, testing hypotheses, and communicating visual insights at a speed and scale beyond human capability. Yet, the emergence of these powerful systems raises critical questions that the VIS community must address: Could autonomous agents eventually replace human data scientists, and if not, how might they best collaborate? Are current visualization techniques and interfaces, originally designed for human analysts, suitable for agent interactions? How can VIS designers effectively integrate agents into their workflows without compromising human agency? And to what extent should agents help shape and educate the next generation of visualization researchers?
Through a mix of keynote talks, paper presentations, and an agentic VIS challenge,
this workshop invites researchers and practitioners to share innovative ideas, explore these 
questions, and discuss strategies to transform the impact of VIS for a future where human and AI agents co-exist.
} 

\nocopyrightspace

\newcommand{\para}[1]{\vspace{1mm}\noindent\textbf{#1}}

\begin{document}


\firstsection{Introduction}

\maketitle

Recent breakthroughs in Artificial Intelligence (AI) are increasingly driven by AI agents, i.e., goal-oriented systems or tools that are capable of iteratively observing, acting upon, and learning from complex environments. Unlike traditional AI models that passively respond to inputs, these agentic systems actively pursue objectives, propose action plans, dynamically adapt, and refine their strategies based on experience. 
For instance, OpenClaw's agent~\cite{openclaw} can autonomously execute multi-step workflows across a user's apps and devices, e.g., triaging an inbox, drafting and sending emails, updating calendars, and driving a browser to search the web.
Tasks that have traditionally required significant human effort can now be accelerated dramatically, illustrating the transformative potential of agents across domains such as scientific discovery, healthcare diagnostics, finance, and environmental monitoring.

For the visualization (VIS) community, the rise of AI agents presents exciting new opportunities to reshape data-intensive workflows. Agents can augment traditional VIS tools by proactively recommending visualizations, interacting with them, identifying overlooked patterns, and even generating complete designs based on data and user needs. Specialized foundation models trained on visualization tasks may further support processes such as visual encoding selection, narrative construction, or interface customization. Looking ahead, agentic systems could enable more advanced scenarios, such as teams of collaborative agents exploring vast visualization parameter spaces or optimizing visual outputs through simulated human feedback. These developments signal a potential shift in how visualizations are designed, evaluated, and deployed, opening the door to more scalable and intelligent visual analytics.

At the same time, these opportunities come with pressing challenges. How can we measure the performance and effectiveness of agents in complex VIS tasks, and what benchmarks or evaluation methods are most appropriate? What ethical, privacy, and security concerns arise when deploying agents in sensitive, data-rich contexts? As embodied and immersive agents become more prevalent, how can we design interfaces that support seamless human–agent collaboration? There are also broader concerns about the long-term role of agents: Will they enhance human expertise, or displace it? And as agents begin to influence education and training, how do we ensure their guidance fosters creativity, critical thinking, and ethical awareness in future visualization practitioners and researchers?

To address these questions and opportunities, 
we launched this community effort last year by organizing the first \textbf{workshop on GenAI, Agents, and the Future of VIS at IEEE VIS 2025}. 
The response was strong: receiving more than 40 submissions and drawing a full-house audience, underscoring both the urgency and excitement around agentic systems and VIS. 
Building on this momentum, this year's workshop invites interdisciplinary collaboration among visualization researchers, AI developers, practitioners, and industry stakeholders. 
Through a mix of keynote talks, paper presentations, and a challenge where participants build and automatically evaluate agentic systems for VIS, we aim to foster meaningful engagement with both the technical and human-centered dimensions of agent integration. By drawing on real-world case studies across healthcare, finance, and environmental monitoring, participants will share insights, identify best practices, and propose strategies for responsibly integrating autonomous agents into visualization workflows. We invite the community to join us in shaping a forward-looking vision for VIS, where humans and AI agents work together to unlock powerful new possibilities in data exploration, analysis, and communication.

\section{Who, Why, What, and How}
\para{Who is this workshop for?} 
This workshop is designed for a diverse set of researchers and practitioners across the broader VIS, human-centered AI, and human–computer interaction (HCI) communities who are interested in the evolving role of AI agents in visualization.
The target audience includes but is not limited to designers and developers building analytic and communication tools, researchers examining human-agent collaboration dynamics, and domain-specific professionals from industry, government, and non-profit sectors integrating AI agents into real-world applications and decision-making.




\para{Why is this workshop needed and different from similar activities?}
This workshop explores the opportunities and challenges of AI agents driven by large foundation models. While workshops such as VISxAI~\cite{visxai} emphasize \emph{visualization for AI explainability}, we focus on \emph{agentic} systems as \emph{actors in the visualization workflow}. 
This framing raises distinct questions about human--agent collaboration, appropriate autonomy, and the accountability and reliability of agent-produced analytical outcomes, complementing (rather than duplicating) existing explainability-focused efforts.
As a research community, we must critically examine both their benefits and potential risks. This workshop will facilitate the exchange of ideas on defining the role of AI agents in data visualization and identifying future research challenges for the community.

\para{What will be discussed?} This workshop will explore how autonomous agents can augment human-centered visualization workflows, covering topics such as building foundational models and agent frameworks for data visualizations, designing novel tools and systems for both agents and humans, developing evaluation metrics and benchmarks for visualization tasks, and addressing ethical, privacy, and security challenges. See Sec~\ref{sec:topics} for more details.

\para{How will this workshop be structured and what is new?} 
This half-day workshop will feature invited speaker(s) presenting topics related to AI agents and human-AI interaction. 
We will accept paper submissions on topics outlined in Sec.~\ref{sec:topics}.
In addition, we will introduce a challenge where participants use AI agents to automatically tackle a data analysis and communication problem. 
Building upon the success of our challenge at VIS 2025~\cite{visxgenai}, 
we plan to expand the scope and rigor of the competition (See Sec~\ref{sec:challenge}).
Accepted paper and challenge submissions will be invited to present their work.

\section{Goals and Topics}

\subsection{Goals and Success Measurement}

This workshop aims to explore the emerging role of AI agents in data visualization by bringing together researchers from visualization, AI, and HCI communities. While recent VIS research has begun to engage with large language models, the broader concept of AI agents remains underexplored. 
This workshop has three primary goals:

\begin{itemize}[leftmargin=* ,itemsep=0pt, topsep=0pt]

    \item  \textit{Goal 1: Define Key Research Challenges}. Identify the unique research opportunities that agentic systems introduce for visualization, such as agent-facing interfaces, agent-driven visual analysis, and human-agent collaboration.

    \item \textit{Goal 2: Bridge Disciplinary Perspectives.} Foster mutual understanding between VIS researchers and the broader AI/agent communities, who may have differing assumptions around goals, evaluation, and system design. 

    \item \textit{Goal 3: Build Automated Infrastructure for Progress.} Explore and initiate an automated evaluation and challenge framework, similar to those successfully adopted in the AI community, to accelerate VIS research through reusable tools, standardized benchmarks, and community participation, beginning with an agentic VIS challenge at this workshop.

\end{itemize}

\para{Evaluating Success and Backup Strategy.}
We define success along three dimensions: intellectual contribution, community engagement, and long-term impact. 
We will track the number and quality of paper and challenge submissions, participant attendance (targeting 50–80), activity on Discord and the playground, and most importantly, the follow-up use and standardization of the challenge framework within the VIS community.

If submissions fall short (e.g., fewer than six), our Boston-based organizers will help recruit local AI and non-VIS participants to adapt relevant work. We will also shift to a discussion- and panel-focused format with expert talks and interactive demos, avoiding the need for emergency submissions.

\subsection{Topics}
\label{sec:topics}
Topics include but not limited to:

\begin{itemize}[leftmargin=* ,itemsep=0pt, topsep=0pt]
    \item \textit{Agent-Augmented VIS Tools for Humans}: How can agents best augment traditional VIS tools to support humans?

    \item \textit{Inventing VIS Tools for Agents:} What new VIS techniques and interfaces do we need to develop for agents to use?

    \item \textit{VIS-Specific Foundation Models and Agents:} How can we design VIS-focused foundation models and specialized agents to revolutionize VIS tasks?

    \item \textit{Collaborative Agents for VIS:} How can multiple collaborative agents transform the way we explore, analyze, and communicate datasets?

    \item \textit{Automated VIS Testing and Feedback through Simulated Humans:} Can synthetic human feedback from agents effectively automate VIS testing and design validation?

    \item \textit{Evaluation Metrics and Benchmarks:} What methods and metrics can we use to evaluate the performance, reliability, and accuracy of agents in VIS tasks?

    \item \textit{Agents Beyond the Desktop:} How can we develop immersive, situated visualizations to facilitate the collaboration between human and embodied agents or robots?

    \item \textit{AI agents in VIS Education:} How can agents be leveraged to educate and mentor the next generation of VIS researchers, fostering creativity, critical thinking, and ethical awareness in data visualization?

    \item \textit{Data Privacy and Security Considerations:} What ethical, privacy, and security challenges arise when deploying agents in VIS, and how can we address them?

    \item \textit{Case Studies and Applications:} How have agents been successfully applied in domains like healthcare, finance, and more, and what can we learn from these cases?

    \item \textit{Future Roles in Data Science:} Will agents augment human capabilities in data science, or could they eventually replace data scientists entirely?
    
    \item \textit{Positioning Papers and Community Building:} How can we foster interdisciplinary collaboration among researchers, practitioners, and industry stakeholders to shape the future of agents in VIS?
\end{itemize}

\section{Pre-workshop Plans}
The workshop will feature invited speakers, paper submissions and presentations, and an agentic VIS challenge (inspired by ImageNet), where participants submit agents to analyze given datasets and generate reports. 

\subsection{Speakers Invitation}
We plan to invite one to two keynote speakers with expertise in human-centered AI to the workshop, ideally one from academia and one from industry (e.g., DeepMind). 
We intend to finalize the list of invited speakers by September 10.

\subsection{Paper Submissions and Presentations}
The workshop welcomes short papers of 2--4 pages (plus up to 2 page references) using the VGTC Conference Style Template.

\para{Submission:}
The submissions are made through the PCS (Precision Conference Solutions) website. 
Submissions are not anonymous and should include all author names, affiliations, and contact information. 
At least one author of each accepted submission needs to register for the IEEE VIS conference and attend the workshop.

\para{Review Procedure:}
A member of the organizing team will assess the abstracts and assign submissions to workshop organizers and invited reviewers based on topical relevance and expertise.
Each paper receives at least two reviews, each including a score (1–5) and written feedback.
A discussion of all submissions will determine which are accepted, 
based on their novelty, quality, and relevance to the workshop.
All the accepted papers will be invited to give a 10-15 mins oral presentations during the workshop,
give poster presentations, and be available on the workshop's website.

\para{Publication Policy:}
Accepted submissions in this category will be published 
as Short position statements / work in progress notes (similar to posters):
available on the workshop's website, not available at a central digital library, 
included in the downloadable proceedings, not considered archival, but possible to be cited. 
Reuse of the content in a follow-up publication or conference paper will be fully allowed.


\subsection{Agentic VIS Challenges}
\label{sec:challenge}
The challenge invites participants to explore how AI agents can automate data visualization and visual analytics. 
Building upon the success of our challenge at VIS 2025~\cite{visxgenai}, 
we plan to \textbf{expand the scope and rigor of the competition}. 
This includes 
1) a broader set of datasets, 
2) more diverse tasks spanning visualization authoring, interpretation, and generative interface design, 
3) a wider range of tools for agents to call, and 
4) semi-automatic evaluation rubrics to enable more consistent and scalable assessment.
Moreover, in addition to traditional interactive webpage format, 
we encourage participants to embed agents into their outputs, while exploring a new form of agent-embedded mediate for data communication.

\para{Provided Resources:}
We will provide several datasets (e.g., VisPubData~\cite{Isenberg:2017:VMC} and/or domain specific datasets
), alongside a multi-agent workflow template capable of 
inputting and analyzing dataset, 
and producing concise one-page reports.
The template will be built using LangGraph~\cite{langgraph} and hosted on GitHub. It will include LangGraph Studio~\cite{lgstudio}, which enables easy monitoring and management of agent behavior and development. Detailed setup instructions will be provided to help participants get started.
We will host an evaluation server that automatically runs each submitted agent configuration 
(in a specified format, such as a ZIP file or GitHub link), generate the one-page report, and post it on the workshop website.

\para{Task:}
Participants are encouraged to build on the existing template by developing agents with improved reasoning and coordination strategies. 
The objective is to generate more insightful, coherent, and visually compelling analyses of the provided datasets, in the form of an interactive web interface.

\para{Submission Process:}
Participants can iteratively refine and submit their agents to the evaluation server until it closes. 
For the final submission, they will upload to the PCS system 
(1) the challenge-server URL hosting their visualization product and (2) a technical report of up to two pages.

\para{Review Procedure and Awards:}
After the submission server closes, each entry will undergo a two-stage evaluation process. First, submissions are assessed automatically against a predefined set of rubrics covering measurable criteria such as task completion, visual encoding correctness, and output quality. Second, each submission is independently reviewed by 2 to 3 human visualization experts who evaluate aspects requiring domain judgment and contextual reasoning. Final scores are derived by combining both assessment components. Submissions will be evaluated based on:
1) \emph{Agent-generated report}: clarity, coherence, and insightfulness;
2) \emph{Technical report}: explanation of key decisions, challenges faced, and lessons learned.
Accepted submissions will be invited to present their work. We will also reach out to sponsors to provide awards for the winning teams.

\para{Publication Policy:}
Awarded submissions will be published in the same format as the short paper submissions.


\subsection{Statement of the Development}

We will assemble the list of contributors (e.g., keynote speakers and paper authors) through open submissions and targeted invitations. We will prioritize topical relevance and diversity in expertise, background, and perspective. 

To encourage broad and inclusive participation, we will promote the workshop across VIS, HCI, and AI communities using social media (e.g., Twitter, LinkedIn, Slack) and targeted email invitations to researchers whose interests overlap with the workshop's topics. 
Our website will host the call for participation, paper submission instructions, schedules, and updates. Two of our co-organizers are based in Boston and will assist in attracting local participants.

\para{Indented important dates}:
\begin{itemize}[itemsep=0pt, topsep=0pt, parsep=0pt, partopsep=0pt]
    \item Call for participation: May 30, 2026
    \item Paper Submission Deadline: Aug 30, 2026
    \item Author Notification: Sep 15, 2026
    \item Camera-Ready Deadline: Oct 1, 2026
    \item Workshop Day: Nov 9 or 10, To be decided
\end{itemize}

\section{Workshop Schedule}
This workshop will be a half-day in-person session (approximately 3.5 hours, including breaks), with remote presentation options for authors who cannot attend. The in-person format fosters community building and interactions among participants, while remote presentations help reduce visa and travel burdens, enabling a more diverse range of attendees.


\begin{table}[ht]
    \centering
    \begin{tabular}{m{2.5cm}|m{5cm}}
    \toprule
    \multicolumn{1}{c|}{ \textbf{Time} } & \multicolumn{1}{c}{ \textbf{Session} } \\
    \midrule
    \multicolumn{2}{c}{ \textbf{\textit{Before the workshop} }} \\
    \midrule
    one week before & 
    In Discord Channel
    \begin{itemize} [noitemsep,topsep=4pt, nosep, leftmargin=*]
            \item Participants introduce themselves
            \item Share all workshop-related material
        \end{itemize} 
    \\
    \midrule
    \multicolumn{2}{c}{ \textbf{\textit{During the workshop}} } \\
    \midrule
    9:00 - 9:10 am \newline (10 min)  & Introduction of workshop organizers, participants, topics, and goals\\
    \midrule
     9:10 - 10:00 am  \newline (50 min)  & Keynote by the invited speaker \newline (discussion and Q\&A included) \\
    \midrule
     10:30 - 11:00 am  \newline(30 min)  & Paper presentation: \newline challenge winners \newline  (3x10 min, Q\&A included)  \\
     \midrule
    \multicolumn{2}{c}{30 min break } \\
     \midrule
    11:30 - 12:20 pm  \newline (50 min)  & Paper presentation: \newline short papers \newline (5x10 min, Q\&A included) \\
    
    \midrule
    12:20 - 12:30 pm   & Workshop summary \\
    \midrule
    \multicolumn{2}{c}{ \textbf{\textit{After the workshop}} } \\
    \midrule
   within one week &  
    Posting summary and records on website. Initiating follow-up activities.
    \\
    \bottomrule
    \end{tabular}
    \vspace{1mm}
    \caption{Proposed workshop schedule}
    \label{tab:workshop_schedule}
    \vspace{-8mm}
\end{table}

\subsection{Planned Activities}
Table~\ref{tab:workshop_schedule} outlines the planned key activities.
Tentatively, the session will run from 9:00 am–12:10 pm (Boston local time) and is subject to change. 
The workshop will begin with a brief 10-minute introduction, followed by 
a 50-minute keynote talk, 
a 30-minute paper session, and a 60-minute paper session.
A 30-minute break between the two sessions will help facilitate networking and reduce fatigue. 
The workshop will conclude with a short summary and discussion of post-workshop plans. Pre- and post-event engagement will be supported via a Discord and a workshop website to share materials, foster discussion, and coordinate follow-up activities.






\subsection{Technical Plans and Activities Facilitation} 
Although the core workshop activities consist primarily of talks, 
we aim to enhance engagement and interaction through the following platforms:

\para{Zoom and Discord Channel:} 
While the workshop is primarily in-person, we will provide live video conferencing (e.g., Zoom) and maintain a dedicated Discord channel. 
This setup, successfully used in previous VIS conferences, enables remote attendance and asynchronous engagement. 
We will stream the in-person talks on Zoom,
and host discussions on Discord channels, where participants can continue conversations before, during, and after the event.
To encourage discussions and networking in Discord, 
we will encourage participants to introduce themselves one week before the workshop. 
The organizing team will also actively coordinate and participate in the Discord
discussions to support the participant engagement.

\para{Web-based Agent Playground:} 
Web-based agent \emph{playgrounds} are now becoming mainstream (e.g.,~\cite{playground}).
We will open the evaluation server and provide a playground user interface (based on LangGraph Studio~\cite{lgstudio}) on the workshop website. 
Participants can 
1) interactively chat with agents hosted on the server,
2) observe and explore the agents' execution processes through a conversational interface.
We will also utilize existing plugins that integrate agent interactions with the Discord channel, making it easy to share or discuss insights directly within Discord.

\para{In-person Posters:}
We will host around 15 posters in the main conference's regular poster session. This approach extends networking opportunities beyond the half-day workshop itself. Attendees are encouraged to continue discussions during the poster session and the scheduled 30-minute coffee break.

\subsection{Expected Audience}
Based on prior workshop experience, we anticipate 50–80 participants, including presenters. 
Through the web-based playground and Discord interaction, we aim to create a unique workshop experience where \emph{both humans and agents contribute to the discussion}.

\section{Post Workshop Plans and Broader Impact}
First, we will use the workshop website as an archival repository for
all workshop-related content. 
Second, depending on participants preferences, we will set up virtual spaces (e.g., slack teams, mail list) for continued discussions and explore the future opportunities.
Third, the interactive website will also serve as a resource for the VIS community to explore agentic systems, and the AI/data science communities to test agent solutions on real-world data analysis problems. 
Finally, we plan to explore the opportunity to write a STAR report with interested
workshop participants.



\section{Organizers}

\para{Chen Zhu-Tian} is an Assistant Professor in the Department of Computer Science and Engineering at the University of Minnesota. His research explores how to augment human intelligence in everyday life through visual interfaces, mixed reality, and AI. He publishes in top venues such as IEEE VIS, ACM CHI, and ACM UIST, where his work has received one Best Paper Award and four Best Paper Honorable Mentions. He also serves as a program committee member or associate chair for these conferences.

\para{Nam Wook Kim} is an Assistant Professor of Computer Science at Boston College, where he directs the Data Within Reach lab (\href{https://dwr.bc.edu}{dwr.bc.edu}). His research is driven by a vision to make data more accessible, bringing it within reach for broader audiences from diverse backgrounds, including blind and low vision individuals. His work has been recognized with awards from ACM CHI, ACM UIST, and the Kantar Information is Beautiful Awards.

\para{Saeed Boorboor}
is an Assistant Professor at the Department of Computer Science at the University of Illinois Chicago, and is part of the Electronic Visualization Laboratory (EVL). His research aims to design visualization systems that empower domain scientists to explore, analyze, and interact with scientific data using methods of image processing, computer graphics, and AI. He publishes in top venues such as IEEE VIS, IEEE VR, and IEEE ISMAR. He also serves as a program and organizing committee member for these conferences.

\para{Shivam Raval} is a final-year PhD student in the Insight and Interaction lab at Harvard, where he works on interpreting and visualizing, and explaining high-dimensional data. Shivam has a background in experimental quantum physics and visualization. His work spans across many disciplines, including experimental physics, data visualization, human-computer interaction, and mechanistic interpretability. Most recently, he was inducted into Sigma Xi, the Scientific Research Honor Society, for his contributions to scientific research.

\para{Pan Hao} is a 2nd year PhD student at the Department of Computer Science and Engineering at the University of Minnesota. His research focuses on enabling AI agents to communicate, coordinate, and act within real-world environments, and designing novel user experiences to enhance human-AI collaboration.

\para{Qianwen Wang}
is an Assistant Professor at the Department of Computer Science and Engineering at the University of Minnesota. Her research aims to enhance communication and collaboration between domain users and AI through interactive visualizations.
Her research has been recognized with awards and featured in prestigious outlets such as MIT News and Nature Technology Features. She has earned multiple recognitions, including two best abstracts from BioVis ISMB, one best paper from IMLH@ICML, two best paper honorable mention from IEEE VIS. 

\para{Vidya Setlur} is the Senior Director of Tableau Research, where she leads an interdisciplinary team of research scientists spanning data visualization, multimodal interaction, applied ML, and NLP. Her core work centers around conversational interfaces for data analytics and led the development of \emph{Ask Data}, Tableau’s natural language question-answering system for visual analytics. Her research draws on methods from information retrieval, human perception, and cognitive science, enabling users to interact with data through intelligent interfaces that bridge human intent and computational reasoning. 

\bibliographystyle{abbrv-doi}

\bibliography{template}
\end{document}